\documentclass[fleqn,10pt]{wlscirep}
\usepackage[utf8]{inputenc}
\usepackage[T1]{fontenc}
\title{Probing non-perturbative QED with electron-laser collisions}

\author[1,*]{C. Baumann}
\author[2,3,+]{E.~N. Nerush}
\author[1]{A. Pukhov}
\author[2,3]{I.~Yu. Kostyukov}
\affil[1]{Institut f\"ur Theoretische Physik I, Heinrich-Heine-Universit\"at D\"usseldorf, 40225 D\"usseldorf, Germany}
\affil[2]{Institute of Applied Physics, Russian Academy of Sciences, 603950 Nizhny Novgorod, Russia}
\affil[3]{Lobachevsky State University of Nizhny Novgorod, 603950 Nizhny Novgorod, Russia}

\affil[*]{Christoph.Baumann@tp1.uni-duesseldorf.de}

\affil[+]{nerush@appl.sci-nnov.ru}

\begin{abstract}
    The vast majority of QED results are obtained in relatively
    weak fields and so in the framework of perturbation theory. However, forthcoming laser facilities
    providing extremely high fields can be used to enter not-yet-studied regimes. Here, a scheme is
    proposed that might be used to reach a supercritical regime of radiation reaction or even the fully 
    non-perturbative regime of quantum electrodynamics.
    The scheme considers the collision of a 100 GeV-class electron beam with a counterpropagating
    ultraintense electromagnetic pulse. To reach these supercritical regimes, it is unavoidable to
    use a pulse with ultrashort duration. Using two-dimensional particle-in-cell simulations, it
    is therefore shown how one can convert a next-generation optical laser to an ultraintense
    ($I\approx 2.9\times 10^{24} \text{ W} \, \text{cm}^{-2}$) attosecond (duration $\approx$ 150 as) pulse.
    It is shown that if the perturbation theory persists in extremely fields, the spectrum
    of secondary particles can be found semi-analytically. In contrast, a comparison with  
    experimental data may allow differentiating the contribution of high-order radiative corrections if the 
    perturbation theory breaks.
\end{abstract}
\begin{document}

\flushbottom
\maketitle
\thispagestyle{empty}

\section*{Introduction}\label{Sec.: I}

The invention of chirped pulse amplification \cite{OptCommun_55_447} almost 30 years ago has led to ever stronger laser systems. State-of-the-art systems are already able to deliver peak intensities of the order of $10^{22}$ Wcm$^{-2}\ \ $\cite{OptExpress_16_2109, HighPowerLaserSciEng_3_e5, OptExpress_25_20486} and future projects aim at reaching even higher intensities \cite{ELI, XCELS, JPhysConfSer_244_032006}. Consequently, the perspectives are opened to explore new regimes of laser-matter interactions. Essentially, the interaction of an electromagnetic (EM) field with an electron can be characterized by two Lorentz invariant quantities \cite{RevModPhys_84_1177}: (i) the dimensionless vector potential $a_0=\vert e \vert \sqrt{-A_{\mu}A^{\mu}}/(m_ec)$ and (ii) the quantum parameter $\chi=\sqrt{-(F_{\mu \nu} p^{\nu})^2}/(m_ecE_{\text{crit}})$. Here, $e$ and $m_e$ are the electron charge and mass, respectively, $c$ is the speed of light, $A_{\mu}$ is the four-vector potential, $F_{\mu \nu}$ is the EM field tensor, $p^{\mu}$ is the four-momentum and $E_{\text{crit}}$ is the critical field of quantum electrodynamics (QED), $E_{\text{crit}}\approx 1.3 \times 10^{16}$ Vcm$^{-1}\ \ $\cite{ZPhys_69_742, PhysRev_82_664}. Simply said, $a_0$ can be seen as a measure for relativistic effects (here $a_0\gg 1$), while $\chi$ accounts for the impact of quantum effects.  Especially the investigation of strong-field quantum effects, like the recoil due to emitted photons (radiation reaction) and the pair production according to the Breit-Wheeler process \cite{PhysRev_46_1087}, is of fundamental interest. These regimes become important when $\chi$ approaches unity and with nowadays experiments one can already probe the regime $\chi \lesssim 1$ with optical fields \cite{PhysRevLett_76_3116, PhysRevLett_79_1626, PhysRevX_8_011020, PhysRevX_8_031004} or with the field of crystals~\cite{NatCommun_9_795} which even allows one to reach $\chi \approx 7$~\cite{PhysRevLett_87_054801}. However, although not yet realizable in laboratories, it is still important to study situations in which $\chi\ggg 1$, referred to as "supercritical regime" \cite{NewJPhys_21_053040} in the following. Such systems are physically relevant, for example for the understanding of extreme astrophysical environments \cite{RepProgPhys_69_2631, RepProgPhys_77_036902, AstroPhysJ_851_129}.  An even more extreme regime is $\alpha \chi^{2/3}\gtrsim 1$ ($\alpha \approx 1/137$ is the fine-structure constant) that has long been assumed to be not accessible experimentally. In this regime, it is believed that quantum processes will be dominated by radiative corrections. This means the emission of virtual photons by an electron or the virtual conversion of a photon in an $e^-e^+$ pair cannot be treated as a small perturbation and in this sense QED becomes a fully non-perturbative theory \cite{AnnPhys_69_555, PhysRevD_21_1176, JPhysConfSer_826_012027}. So far, it does not exist a closed theory describing this highly nonlinear regime, and there are only a couple of works that have studied this regime over the last decades~\cite{JPhysConfSer_826_012027,PhysRevD_99_076004,PhysRevD_99_085002}.

This manuscript shows that it might be possible to probe the regime $\alpha \chi^{2/3}\gtrsim 1$ in an electron-EM pulse collider setup in the future. Considering an ultrarelativistic electron, the quantum parameter can be written approximately as $\chi_e \approx \gamma_e\, F_{\perp}/(|e|E_{\text{crit}})$, where $\gamma_e$ is the Lorentz factor of the electron and $F_{\perp}$ is the force acting perpendicular to its velocity. In terms of the quantum parameter $\chi_e$, the fully non-perturbative QED regime sets in at $\chi_e\sim 1600$.  Assuming an electron with $\gamma_e=2.5\times10^5$ ($\approx$ 125 GeV energy), a field strength of $8.3\times 10^{13}$ Vcm$^{-1}$ is required to reach this completely novel regime of QED.  In principle, the next generation of laser systems will deliver such extreme field conditions \cite{ELI,XCELS,JPhysConfSer_244_032006}. However, discussing the characteristic time between two photon emission events in the case $\chi \gg 1$~\cite{QuantumElectrodynamics, PhysRevSTAB_14_054401}, 
\begin{align}
\label{Eq.: 1}
\begin{split}
t_{\text{rad}} \sim W^{-1}_{\text{rad}} = \left(1.46 \,\frac{m_ec^2}{\hbar \gamma_e}\, \alpha \chi_e^{2/3}\right)^{-1} \sim 200 \text{ as},
\end{split}
\end{align}
one can see that for optical lasers, even a single-cycle pulse, which duration is $\sim 3 \text{ fs}$, is not sufficiently short ($\hbar$ is Planck's constant). Electrons will radiate almost their entire energy in terms of high-energetic $\gamma$-photons before they reach the region of maximal field strength. Consequently, getting access to the regime $\alpha \chi_e^{2/3}\gtrsim 1$ is not easily achieved with conventional laser pulses of ultra-high intensity.  Therefore, it is unavoidable to develop experimental schemes that mitigate these radiative losses significantly.  This can be done, for instance, by reducing the interaction time between electrons and the background field. Here, a first approach has been presented by Yakimenko $et\ al.$ \cite{PhysRevLett_122_190404}. Instead of using strong lasers sources, that work suggests to use a future 100 GeV electron-electron collider. To achieve the required field strengths, the electron beams are supposed to be extremely dense and tightly focused transversely ($\sigma_\perp =10$ nm), leading to a strong peak current ($I_{\text{max}}=1.7$ MA) and so to strong collective self-fields. Mitigation of radiative losses is ensured by a strong longitudinal compression of the beams ($\sigma_\parallel =10$ nm). Another approach how one could probe the impact of radiative corrections has been proposed by Blackburn and coworkers \cite{NewJPhys_21_053040}. They raised the idea of colliding highly energetic electrons with ultraintense laser pulses. To mitigate radiative losses of the electrons, they put an oblique scattering geometry forward. In doing so, it seems possible to reach $\chi_e$-values of the order of 100. The present manuscript, however, aims at introducing a setup that makes a supercritical regime of radiation reaction and pair production accessible, and that may even be promising for exceeding the $\chi_e$-threshold of $1600$. In particular, it is discussed how it may become possible to enter the regime $\alpha \chi_e^{2/3}\simeq 1$ even in the head-on collision of a high-intensity pulse with a 125 GeV electron beam. Therefore, it is shown how one can convert a next-generation optical laser pulse to an extremely intense attosecond pulse. The observed attosecond duration ($\sim$ 150 as) is sufficiently reduced to suppress radiative losses significantly. Based on the Ritus--Nikishov theory for strong-field QED processes \cite{SovPhysJETP_25_1135}, numerical as well as analytical calculations are presented to investigate the interaction of ultrarelativistic beam electrons with extremely intense fields. The numerical simulations underline the feasibility of the proposed scheme.

\section*{Results}
\subsection*{Proposed scheme and its numerical modeling}\label{Sec.: II}
\begin{figure}[htb]
\centering
\includegraphics{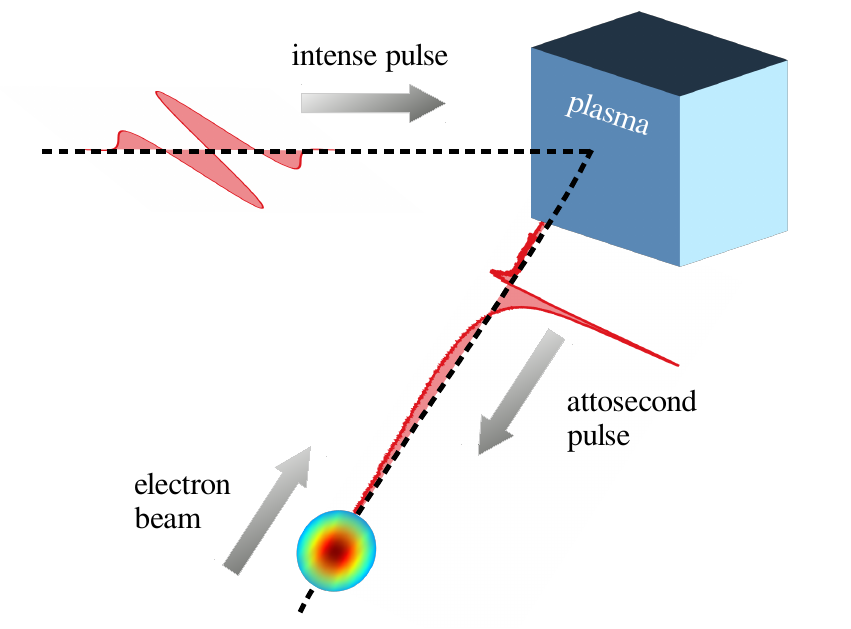}
\caption{The figure shows a sketch of the proposed scheme. A moderately next-generation laser pulse impinges at oblique incidence ($\theta=30^\circ$) onto a overdense plasma target ($150\;n_{\text{cr}}$). The plasma converts the incoming radiation into an ultraintense attosecond pulse via high harmonic generation.}
\label{Fig.: 1}
\end{figure}
To corroborate the possibility of reaching the fully non-perturbative QED regime in the collision
of an electron beam with an ultrashort EM pulse, two-dimensional particle-in-cell (PIC) simulations
are performed with the code VLPL \cite{JPlasmaPhys_61_425,CernYellRep_1_181}. The PIC code can be
used to incorporate QED events into the self-consistent modeling of the laser-particle interaction. In the
simulations, it is therefore possible to account for the emission of $\gamma$-photons by
ultrarelativistic particles via the nonlinear Compton scattering process and for the creation of
electron-positron pairs according to the multiphoton Breit-Wheeler process. These processes are
included in terms of the widely used Monte-Carlo approach \cite{PlasmaPhysControlFusion_51_085008,
PhysRevSTAB_14_054401, PhysRevE_92_023305}. A benchmark of the QED module that is implemented into
VLPL against a similar code can be found in \cite{PhysRevE_94_063204}. Note that the QED module is
based on the Ritus--Nikishov formulas which do not account for high-order radiative corrections so
far (probably needed at $\alpha \chi_e^{2/3} \gtrsim 1$). Consequently, it may be that the
method models particles in such extreme states only inappropriately. The
Ritus--Nikishov results, however, are far apart from being useless. Performing the proposed
experiment and comparing the experimental data with the results of the Ritus--Nikishov theory (see
the discussion in the following sections) allows important insights in at least two unsolved questions:
\begin{enumerate}[label=(\roman*)]
		\item Is the Ritus--Nikishov theory applicable in this regime?
        \item If the Ritus--Nikishov theory is inapplicable in this regime, how will the observables be modified and what can be learned 				  from that?
\end{enumerate}
Thus, the approach presented here is a logical step towards a better understanding of superstrong-field QED.

As already pointed out in the introduction, it is necessary to convert the laser radiation into an attosecond pulse with ultra-high intensity. To fulfill this requirement, the proposed scheme follows the approaches previously presented in works considering the generation of high harmonics \cite{PhysPlasmas_17_033110, PhysRevE_84_046403}. A p-polarized single-cycle laser pulse (central wavelength $\lambda_0=1\mu$m) impinges at oblique incidence onto an over-dense plasma slab (see sketch in Fig.~\ref{Fig.: 1}). In principle, a generalization to longer laser pulses is possible when using the attosecond lighthouse technique~\cite{PhysRevLett_108_113904}. The laser has a Gaussian shape in the transverse direction and it is focused to a spot size of $w_0=2.5\lambda_0$ at the focal point $x_0=10\lambda_0,\ y_0=0$. The dimensionless vector potential $a_0$ is set to 350, which corresponds to a peak intensity of $1.68\times10^{23}\,$Wcm$^{-2}$ and to a total power of $35$ PW. The angle of incidence, measured with respect to the target's normal direction, is equal to $\theta=30^\circ$. The target is modeled by a slab with initial density $150n_{\text{cr}}$, where $n_{\text{cr}}=1.12\times10^{21}$cm$^{-3}$ is the critical density for the above wavelength. It is assumed to be fully ionized initially. The ion mass-to-charge ratio is chosen to be two times that of protons, which means that one could potentially use any fully ionized low-Z species as a target material. It is also important to stress that the plasma slab is attached to a plasma density ramp, modeled as $n\propto\exp\left[(x-x_0)/(0.33\lambda_0)\right]$ for $x<x_0$. The plasma profile impacts the efficiency of the generation of high harmonics and defines the requirements for the contrast of the high-power laser pulse~\cite{PhysRevLett_109_125002}. The probing electron beam propagates under the same angle $\theta$ such that it can hit the reflected radiation in a nearly head-on scenario. The beam profile is Gaussian-like with rms length $\sigma_{\parallel}=\lambda_0/40$ and rms width $\sigma_{\perp}=\lambda_0/5$. Each beam electron has an initial $\gamma_e$-factor of $2.5\times10^5$, corresponding to an energy of roughly 125 GeV. The beam density is determined such that the beam yields a peak current of $I_{\text{max}}\approx13.5$ kA and a total charge of $Q\approx 2.8$ pC. Initially, the beam is shifted spatially so that it hits the EM attosecond pulse in its focus. The simulation box has a size of $15\lambda_0$ in the $x$ direction and $20\lambda_0$ in the $y$ direction with a cell size of $\Delta x=\Delta y= 0.005\lambda_0$. 

\subsection*{Simulation results}\label{Sec.: III}
\begin{figure}[htb]
\centering
\includegraphics{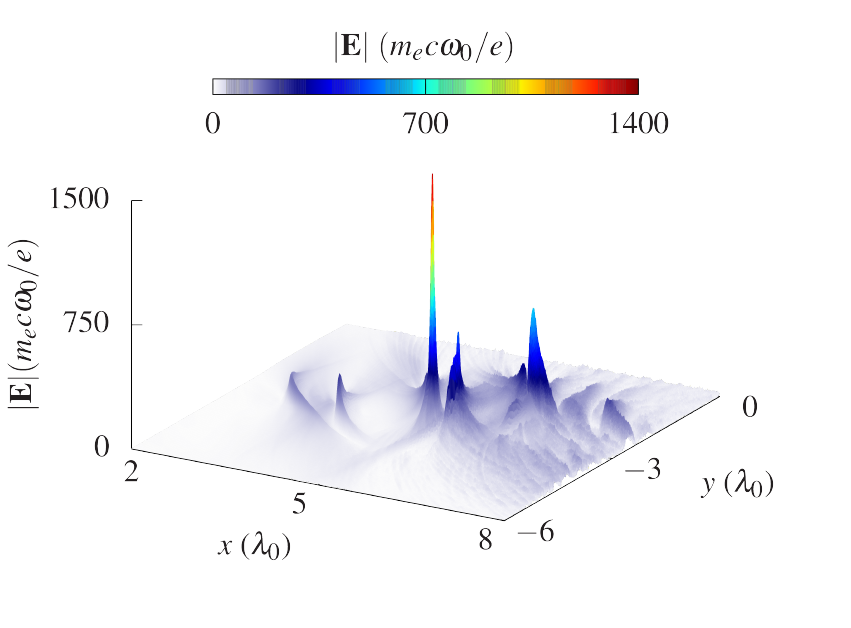}
\caption{The plot illustrates the generated light pulse in terms of the absolute value of the electric field at time $t=13T_0$. At this moment in time the pulse reaches its peak value.}
\label{Fig.: 2}
\end{figure}

\begin{figure}[htb]
\centering
\includegraphics{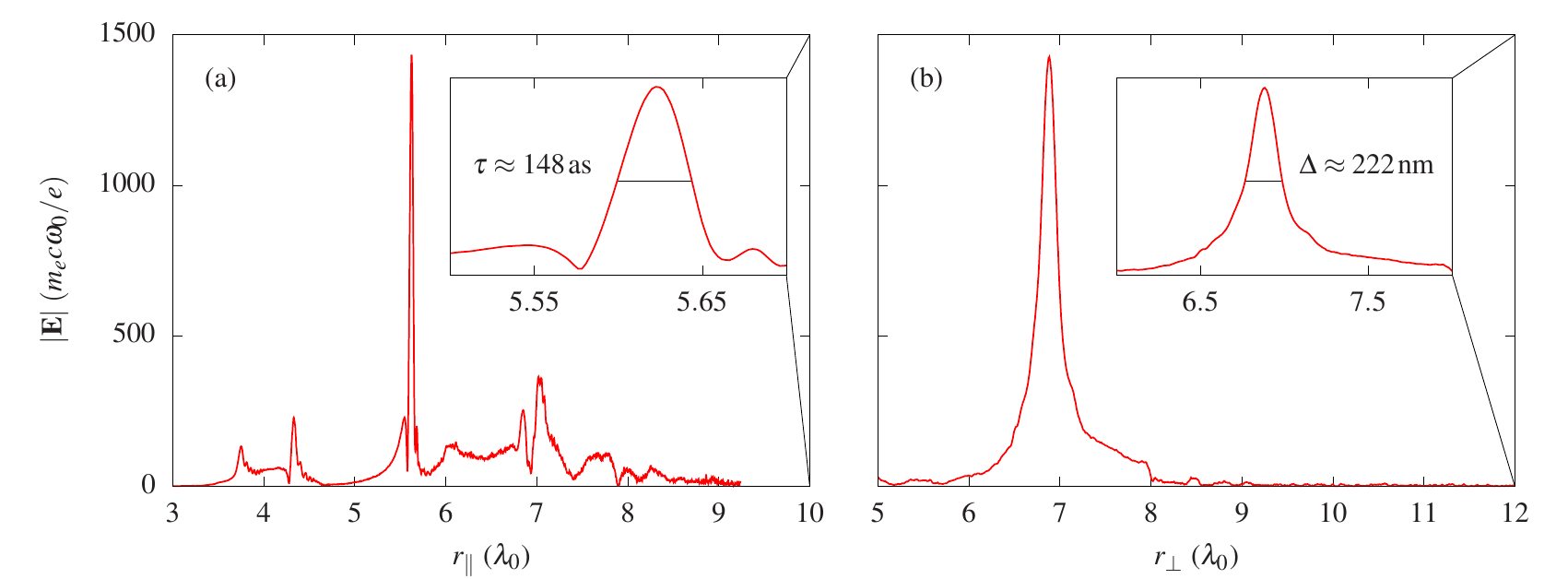}
\caption{The figure shows a characterization of the (a) longitudinal profile of the generated EM pulse, seen by the ultrarelativistic electron beam. The inset shows a zoom of the attosecond pulse, with a FWHM duration of $\approx$ 148 as. (b) The transverse profile of the attosecond pulse at the focus is shown, indicating a FWHM width of $\approx$ 222 nm (inset).}
\label{Fig.: 3}
\end{figure}

Figure~\ref{Fig.: 2} presents the results for the reflected radiation in terms of the absolute value of the electric field at time $t=13T_0$. At this time instant, the EM field is most tightly focused in the transverse direction and it is most compressed in the longitudinal direction, leading finally to the emergence of an ultraintense and ultrashort EM pulse. A transverse and a longitudinal characterization of the pulse is given in Figs.~\ref{Fig.: 3}(a) and (b), respectively. In particular, Fig.~\ref{Fig.: 3}(a) shows a cut of $\vert \mathbf{E}\vert$ along the propagation axis of the electron beam (angle $\theta=30^\circ$).  The length $r_{\parallel}$ stands here for the distance of a point on the propagation axis with respect to the intersection point with the $y$-axis. It can be seen that the main peak of the electric field protrudes from the rest of the field structure and one finds a maximum value of $\vert \mathbf{E}\vert \approx 1450$. In addition, the focused pulse appears to be very narrow on the applied length scale. Therefore, the inset shows an enlarged picture of the main peak. One can extract a duration (full width at half maximum, FWHM) of roughly 150 as from the data. Comparing this with the radiation time in Eq.~\eqref{Eq.: 1} ($t_{\text{rad}}\approx $ 200 as), the duration of the attosecond pulse is less than this characteristic time between two photon emission events. Consequently, the attosecond pulse might be sufficiently short to prevent electrons from radiating all their kinetic energy too fast and so making it possible to reach the highest possible values of $\chi$.  Figure~\ref{Fig.: 3}(b) displays a cut of the field profile along the direction perpendicular to the beam axis at $r_\parallel\approx5.625\lambda_0$. It can be seen that, besides its ultrashort duration, the generated pulse is characterized by a tight focal width of approximately 220 nm.

\begin{figure}[htb]
\centering
\includegraphics{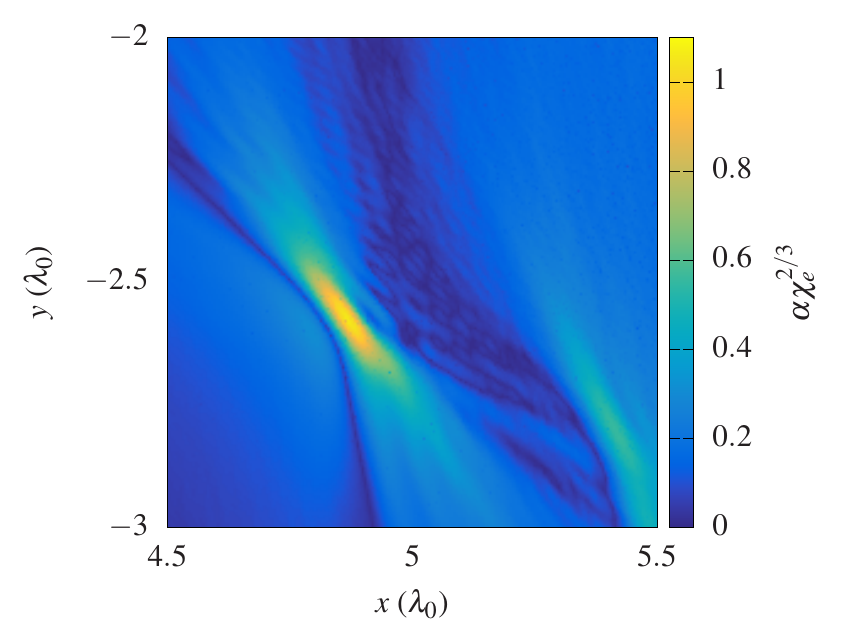}
\caption{Non-perturbative quantum parameter $\alpha\chi_e^{2/3}$ at time $t=13T_0$. The simulation results indicate that the regime $\alpha\chi_e^{2/3}\gtrsim 1$ might be accessible in the proposed scheme.}
\label{Fig.: 4}
\end{figure}

Figure~\ref{Fig.: 4} shows the maximum value for the fully non-perturbative QED parameter $\alpha\chi_e^{2/3}$ in each simulation cell. It can be seen that the electrons in the vicinity of the focal spot indeed experience such extreme conditions that one could probe the breakdown of perturbation theory. Notably, a maximum value of $\chi_e\approx 1750$ can be observed in the simulations.

\subsection*{Model}\label{Sec.: IV}
The following section presents an analytical model for the interaction of ultrarelativistic particles with strong EM fields in the case $\chi\gg1$. The model describes the evolution of the system in a self-consistent manner since it also takes into account the generation of secondary particles. A comparison with particle spectra obtained from simulations indicates reasonable agreement.

The analytical approach is based on the distribution of particles in the phasespace. The corresponding time evolution of the system is described by the Boltzmann equations (BE, for electrons, photons and positrons; see, for example~\cite{PhysRevSTAB_14_054401}).
The electron trajectories are bent by the EM field only by the negligible angle $|e| E \tau /
(m_e c \gamma_e) \sim 10^{-3}$, thus one can neglect the transverse particle dynamics and proceed to a
one-dimensional BE for the particle distribution functions. In this one-dimensional
description the external fields are taken at position $x(t) = x(0) +
ct$~\cite{PhysRevA_87_062110}, and the distribution functions are the same as
the particle spectra. As will be shown later, the exact dependence of the external field
on time does not matter if $\chi \gg$ 1, and the global constant field approximation can be applied
(see also Ref.~\cite{PhysRevLett_87_054801} for more information). Thus, the BE in a constant
homogeneous magnetic field of the strength $H$ can be used instead of using an alternating laser
field. It yields that the particle's phasespace evolution is governed by
\begin{eqnarray}
    \label{f_e}
    \partial_t f_{e,p} = -W_\text{rad} f_{e,p}
    + \int_\varepsilon^\infty f_{e,p}(\varepsilon') w_\text{rad}(\varepsilon' \to \varepsilon) \, d\varepsilon' 
    + \int_\varepsilon^\infty f_\gamma(\varepsilon') w_\text{pp}(\varepsilon' \to \varepsilon) \, d\varepsilon',
\end{eqnarray}
\begin{eqnarray}
    \label{f_gamma}
    \partial_t f_\gamma = -W_\text{pp} f_\gamma 
    + \int_\varepsilon^\infty \left[ f_e(\varepsilon') + f_p(\varepsilon') \right] w_\text{rad}(\varepsilon'
    \to \varepsilon' - \varepsilon) \, d\varepsilon',
\end{eqnarray}
where all distribution functions are given at $(t, \varepsilon)$ unless otherwise specified. Additionally, $w_\text{rad}(\varepsilon'
\to \varepsilon)$ is the differential emission probability rate for the electron whose energy drops from
$\varepsilon'$ to $\varepsilon$ after the emission of a single photon, $w_\text{pp}(\varepsilon' \to
\varepsilon) \equiv w_\text{pp}(\varepsilon' \to \varepsilon' - \varepsilon)$ is the differential
probability rate for a photon with energy $\varepsilon'$ to produce an electron with energy
$\varepsilon$ and a positron with energy $\varepsilon' -
\varepsilon$, and
    $W_{\text{rad},\text{pp}}(\varepsilon) = \int_0^\varepsilon
    w_{\text{rad},\text{pp}}(\varepsilon \to \varepsilon') \, d\varepsilon'$
are the full probability rates for photon radiation and for pair photoproduction, respectively.
Since the emission and pair production differential probabilities at $\alpha \chi^{2/3} \gtrsim  1$ 
are not yet known, the Nikishov--Ritus (Baier--Katkov) probabilities~\cite{QuantumElectrodynamics, SovPhysJETP_25_1135, PhysRevSTAB_14_054401} are used instead in the following.

To further advance the analytical calculations, one first considers energies of the secondary
particles that correspond to $\chi \gg 1$. In this regime, the differential probabilities can be written as~\cite{PhysPlasmas_18_083107}
\begin{eqnarray}
 \label{wapprox}
   w_\text{rad}(\varepsilon' \to \varepsilon) = \frac{\nu}{ {\varepsilon'}^{4/3}} \frac{1 + \eta^2}{\eta^{1/3} (1 - \eta)^{2/3}}    
   \left( \frac{H}{H_\text{crit}} \right)^{2/3},\\
 \label{tildewapprox}
   w_\text{pp}(\varepsilon' \to \varepsilon) = \frac{\nu}{ {\varepsilon'}^{4/3}} \frac{\eta^2 + (1 - \eta)^2}{\eta^{1/3} (1 - \eta)^{1/3}}
   \left( \frac{H}{H_\text{crit}} \right)^{2/3}.
\end{eqnarray}
Here, $\nu$ and $\eta$ are defined by the relations
\begin{eqnarray}
    \nu = -\frac{\alpha \operatorname{Ai}'(0) \left(m_e c^2\right)^{4/3}  }{\hbar}, \quad
    \eta = \frac{\varepsilon}{\varepsilon'},
\end{eqnarray}
with $\operatorname{Ai}^\prime$ being the derivative of the Airy function. The BE is a linear
equation whose solution can be expressed in the general case with time-ordered exponentials~\cite{PlasmaPhysControlFusion_61_074003}, and the
dependence of these differential probabilities on the field strength $H$ suggests the following
assumption. If $H$ is depending on time in two distinct systems, namely $H = H_1(t)$ in one system
and $H = H_2(t)$ in the other,
and $f_{e,p,\gamma}$ are initially ($t=0$) the same for both systems, 
the distribution functions will be also the same at time $t$ for both systems if
\begin{eqnarray}
    \label{correspondance}
 	 \int_0^t H_1^{2/3}(t^\prime) \, dt^\prime = \int_0^t H_2^{2/3}(t^\prime) dt^\prime.
\end{eqnarray}
This $H^{2/3}$-correspondence demonstrates that in the supercritical regime the
spectra of particles with $\chi \gg 1$ can be modelled without the knowledge of the exact
shape of the EM pulse. Also the $H^{2/3}$-correspondence motivates the use of the global constant field 
approximation.

If $t_\text{rad} \lesssim \tau$, the solution of Eqs.~\eqref{f_e} and~\eqref{f_gamma} can be found in the framework of
perturbation theory, 
\begin{eqnarray}
    \label{pert_theory}
 	f(t) = f^{(0)} + f^{(1)} + f^{(2)} + ...\, ,
\end{eqnarray}
where $f^{(i)} \propto (t / t_\text{rad})^i$ describes the $i$-th generation of the secondary particles. 
Substituting the initial energy distributions $f_e^{(0)} = \delta(\varepsilon - \varepsilon_0)$,
$f_{p, \gamma}^{(0)} = 0$ into the right-hand-side of Eqs.~\eqref{f_e} and~\eqref{f_gamma}, one gets
\begin{eqnarray}
    \label{f1e}
    f_e^{(1)} =  t\cdot\left[w_\text{rad}(\varepsilon_0 \to \varepsilon) -W_\text{rad} \delta(\varepsilon - \varepsilon_0) \right], \\
    \label{f1gamma}
    f_\gamma^{(1)} = t \cdot w_\text{rad}(\varepsilon_0 \to \varepsilon_0 - \varepsilon).
\end{eqnarray}
It can be clearly seen from these formulas that from the experimental point of view it is highly
desirable to have a very short EM pulse, $\tau \ll t_\text{rad}$. In this case $f \approx
f^{(1)}$, and the shapes of $f_{e, \gamma}(\varepsilon)$ directly reproduce the shape of
$w_\text{rad}(\varepsilon_0 \to \varepsilon)$. Thus, measuring $f_{e, \gamma}(\varepsilon)$, one can find
$w_\text{rad}(\varepsilon_0 \to \varepsilon)$, including the case $\alpha \chi^{2/3} \gtrsim 1$.

In order to take into account the next terms of the perturbation theory, one considers the interval $\varepsilon_0/\chi_0 \ll \varepsilon \ll \varepsilon_0$, and obtains from Eqs.~\eqref{f1e} and~\eqref{f1gamma} 
    $f_e^{(1)}(\varepsilon) \propto \varepsilon^{-1/3}$,
    $f_\gamma^{(1)}(\varepsilon) \propto \varepsilon^{-2/3}$.
By integrating over $1/\eta$ instead of $\varepsilon'$, it can be easily shown that any convolution from Eqs.~\eqref{f_e} and~\eqref{f_gamma} for a power-law distribution function $f \propto \varepsilon^s$ yields again a power-law distribution with $f \propto \varepsilon^{s-1/3}$.
The same holds for any multiplication by $W_\text{rad}$ or $W_\text{pp}$ in the BE, since $W_\text{rad} \sim W_\text{pp} \propto \varepsilon^{-1/3}$. Thus, substituting $f_e^{(1)}$ and $f_\gamma^{(1)}$ into the
right-hand-side of Eqs.~\eqref{f_e} and~\eqref{f_gamma}, one gets $f_{e,p,\gamma}^{(2)}$, and,
summarizing up to the second order of the perturbation theory on the interval where
$\chi \gg 1$ and $\varepsilon \ll \varepsilon_0$, one finally has
\begin{eqnarray}
    \label{fefit}
    f_e \approx a \varepsilon^{-1/3} + b \varepsilon^{-2/3} + c \varepsilon^{-1}, \\
    \label{fpfit}
    f_p \propto \varepsilon^{-1}, \quad
    f_\gamma \approx g \varepsilon^{-2/3} - h \varepsilon^{-1},
\end{eqnarray}
where $a$, $b$, $c$, $g$ and $h$ are positive quantities.

For the proposed setup the EM pulse duration ($\tau \approx 150 \text{ as}$) is about the radiation 
time ($t_\text{rad}\approx 200 \text{ as}$) and it is also about the characteristic time of pair 
production, so one expects a single generation of secondary particles from the peak of the pulse. 
However, the duration and the amplitude of the post- and prepulses [$\tau \sim 3 \text{ fs}$ and 
$E \sim 100$, respectively, see Figs.~\ref{Fig.: 3}~(a) and~(b)] are such that for the prepulse 
$t_\text{rad} \sim 1 \text{ fs}$ and $\tau / t_\text{rad} \sim 3$, so one expects several 
generations of secondary particles from it.

\begin{figure}
    \centering
    \includegraphics{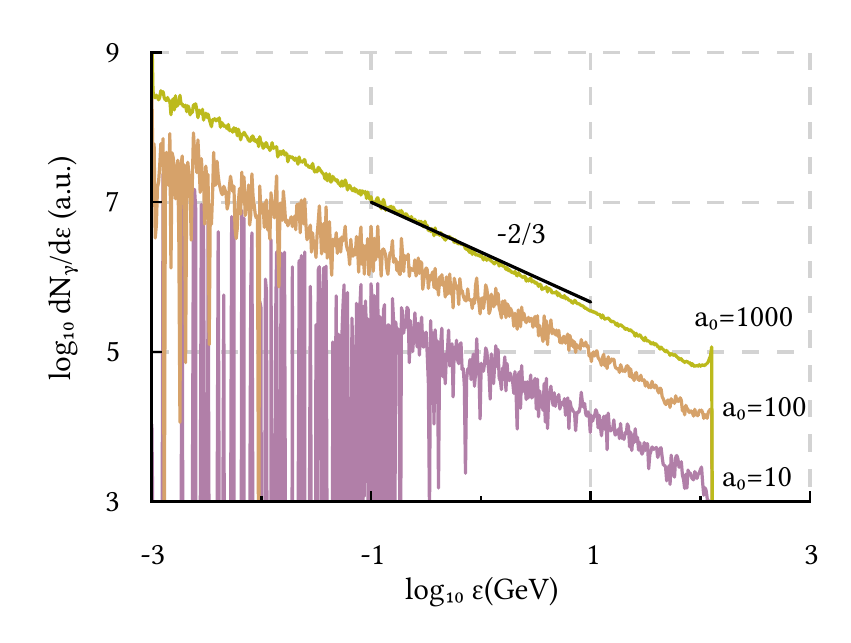}
    \caption{Spectrum of the photons after the interaction of a
    $125 \text{ GeV}$ electron beam with a clean $150 \text{ as}$ Gaussian EM pulse
    of different amplitudes $a_0 = 10$, $100$ and $1000$. The black 
    line shows the function $f \propto \varepsilon^{-2/3}$.}
\label{Fig.: 5}
\end{figure}

\subsection*{Approaching non-perturbative QED}\label{Sec.: V}

Prior to entering the regime of fully non-perturbative QED, experiments will surely approach a perturbative supercritical regime. 
Such a regime is characterized by a large quantum parameter ($\chi \gg 1$), but perturbative QED calculations
are still mostly correct so that the Nikishov--Ritus formulas are applicable. In the following, it is
explained how one can identify the perturbative supercritical regime, and how one probably detect the break of 
perturbative QED, so enabling the determination of its range of validity. Therefore, a series of additional PIC simulations
is performed in a one-dimensional geometry. Note that the attosecond pulse is now assumed to
have a perfect temporal Gaussian profile, $a=a_0\,e^{-(x-ct)^2/\sigma_\tau^2}$, instead of being
generated in a self-consistent way. This kind of pulse is subsequently referred to as a clean
Gaussian pulse. The additional simulations are also relevant in order to test the developed analytical model 
for a few generations of secondary particles, to test the $H^{2/3}$-correspondence, and to figure out the role 
of the prepulse.

If $\chi \gg 1$, the Nikishov--Ritus formulas predict that the shape of the probabilities $w_\text{rad,
pp}$ remains the same for different field strengths, see Eqs.~\eqref{wapprox} and
\eqref{tildewapprox}. Simultaneously, the spectrum of the generated photon beam will reproduce the
shape of $w_\text{rad}$ if the EM pulse is short enough, as shown in the previous section. Thus, the shape of 
the photon spectrum remains the same when one increases the amplitude of an ultrashort EM pulse. This
is in a good agreement with Fig.~\ref{Fig.: 5}, where the photon spectra after the interaction
of the electron beam with clean $150 \text{ as}$ EM pulses of different strengths ($a_0 = 10$, $100$
and $1000$) are given. The electron beam parameters remain unchanged (see the numerical modeling section for more information).
Consequently, maximum values of approximately $12,\ 120$ and $1200$ are expected for $\chi$. One can observe that 
all photon spectra obey the same power-law behavior that coincides with the behavior of the first generation of 
the photons predicted by Eq.~\eqref{fpfit}. Thus, for ultrashort EM pulses, a change of the photon spectrum shape can indicate
the entering in the regime of non-perturbative QED.

\begin{figure}
    \centering
    \includegraphics{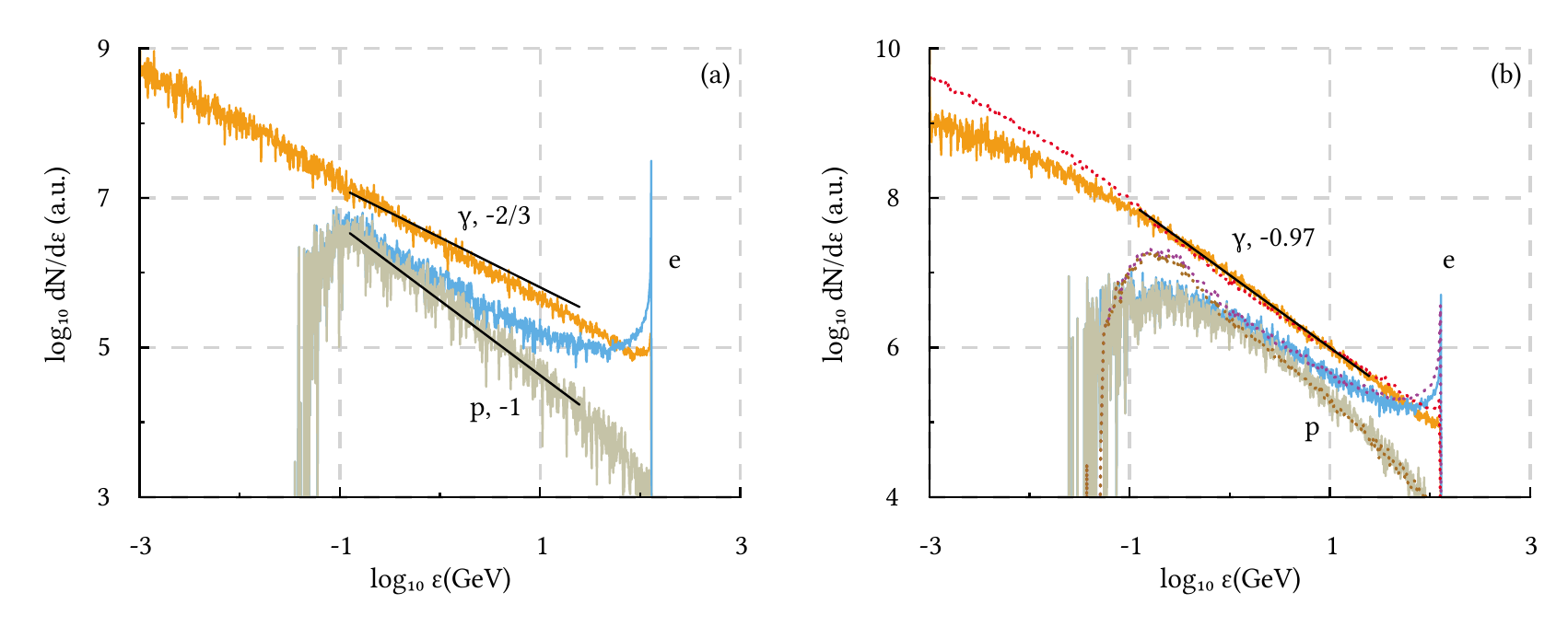}
    \caption{Spectrum of the photons ($\gamma$), electrons (e) and positrons (p) after the interaction of a
    $125 \text{ GeV}$ electron beam with (a) a clean $150 \text{ as}$ Gaussian EM pulse and (b) the EM pulse
    generated in the proposed setup (solid lines) and a clean 350 as Gaussian pulse (dashed lines). Black 
    lines show the fits $f \propto \varepsilon^s$ on the interval from $125 \text{ MeV}$ to $25 \text{ GeV}$, 
    with $s$ given in the labels.}
\label{Fig.: 6}
\end{figure}

To test the developed analytical model for a few generations of secondary particles, the interaction 
of the electron beam with clean attosecond EM pulses of various FWHM durations ($\tau= 50$, $100$, 
$150$ and $350 \text{ as}$) and a fixed dimensionless field amplitude ($a_0=1450$) is investigated in 
the following. Figure~\ref{Fig.: 6}~(a) shows the results for the collision with a clean Gaussian EM 
pulse that has a duration of $150\ \text{as}$. This duration is used since it matches the one of the 
generated main pulse [see Fig.~\ref{Fig.: 3}~(a)], and hence can be later used as a reference.  
Besides, Fig.~\ref{Fig.: 6}~(b) displays the results (solid lines) for the collision with the EM pulse 
that is generated in the laser-plasma interaction described in this manuscript. The additional black 
lines show the power-law fits $f \propto \varepsilon^s$ that are expected to describe the particle 
distribution function according to Eqs.~\eqref{fefit} and~\eqref{fpfit}. The values of the power-law 
indexes are given in the labels. It can be seen in Fig.~\ref{Fig.: 6}~(a) that the power-law
functions~\eqref{fefit} and~\eqref{fpfit} are in reasonable agreement with the simulation results,
indicating the regime of single generation of the secondary photons and positrons. However, note
that the best power-law index for the photon distribution in Fig.~\ref{Fig.: 6}~(a) is
approximately $-0.77$ which is close but not exactly equal to $-2/3$. This may indicate that the
photon distribution function is already influenced by further generations of secondary photons.
Therefore, simulations with shorter pulses are carried out. One obtains best-fitting power-law
indexes of $-0.73$ and $-0.67$ for $\tau = 100 \text{ as}$ and $50 \text{ as}$ laser pulses,
respectively.

If the EM pulse is not short enough, more generations of secondary particles should be taken into
account. This can be seen, for instance, when returning to the photon spectrum that is obtained 
in the proposed setup [see solid lines in Fig.~\ref{Fig.: 6}~(b)]. A deviating power-law index indicates that the analytical
solution~\eqref{fefit} and~\eqref{fpfit} is inapplicable, and hence the pre- and postpulses of the
attosecond pulse have to be included in the BE, e.g. by solving it numerically. However,
the $H^{2/3}$-correspondence allows one to reproduce the spectra of particles with $\chi \gg 1$
($\varepsilon \gg 100 \text{ MeV}$) even in simple one-dimensional simulations, and so still enables a 
comparison of the experimental results and the predictions of perturbative QED. This can be
achieved as follows.  First, one has to retrieve the time integral over $H^{2/3}$ along the
electron trajectories for the attosecond pulse generated in the laser-plasma interaction. Then, one
can find the duration of a clean Gaussian pulse that yields the same value of the time
integral over $H^{2/3}$. For the self-consistently generated attosecond pulse, this procedure leads
to a duration of $350 \text{ as}$. The interaction of such a pulse with the electron beam results
in particle spectra that coincide very good with the spectra obtained in the proposed setup, as can
be seen from the dashed curves in Fig.~\ref{Fig.: 6}~(b). The comparison with the reference spectra
in Fig.~\ref{Fig.: 6}~(a) finally gives insights into the impact of the prepulse on the
distribution of high $\chi$ particles ($\chi \gg 1$, $\varepsilon \gg 100 \text{ MeV}$).
Alternatively, the Gaussian pulse duration can be chosen to ensure this spectrum coincidence, for
instance, when the time integral over $H^{2/3}$ is not known for the generated attosecond pulse. 

Therefore, if the duration of the EM pulse is such that a few generations of secondary particles arise, 
one can use the $H^{2/3}$-correspondence to find the particle spectra in the framework of perturbative QED. 
Thus, variations in the emission and pair production probabilities caused by high-order radiative corrections 
will be detectable if they either disturb the $H^{2/3}$-correspondence or the Ritus--Nikishov formulas.

\section*{Conclusions}\label{Sec.: VI}
In conclusion, an experimental scheme has been proposed that might be promising for exploring
a supercritical regime of radiation reaction or even a fully non-perturbative regime of QED. 
This scheme considers the generation of an ultraintense attosecond pulse via high harmonic 
generation at an over-dense plasma surface. In a second step, the EM attosecond pulse collides with 
a counterpropagating 100 GeV-class electron beam in a nearly head-on geometry. Numerical PIC simulations
underline the feasibility of the proposal.

An analytical treatment reveals that if the attosecond pulse is short enough, the number of 
secondary particles will be small and the differential probability of the photon emission can be 
measured directly in the proposed scheme. 
In this regime the Ritus--Nikishov differential probabilities yield power-law spectra
for the resulting photon and positron beams, $f_\gamma \sim \varepsilon^{-2/3}$ and $f_p \sim
\varepsilon^{-1}$.
Based on perturbative QED, the model further indicates that in the case of
relatively long attosecond pulses the most part of the resulting particle spectra can be reproduced 
in simple one-dimensional simulations with clean Gaussian pulses of appropriate duration. 
If the Nikishov--Ritus formulas break, simple one-dimensional simulations will not be able to reproduce the 
experimental spectra any more, indicating the entering of the regime where high-order radiative corrections 
become important.

\section*{Data Availability}
The datasets generated during and/or analyzed during the current study are available from the 
corresponding author on reasonable request.

\section*{Acknowledgements}

This work has been supported in parts by the EUROfusion Consortium under project No AWP17-ENR-MFE-FZJ-03, 
by the German Research Foundation (DFG) under the project PU 213-6/1 and by the German BMBF under project 
No 05K2016. The solution of the Boltzmann equation has been developed with support of the Russian Science 
Foundation through Grant No. 18-11-00210. We also acknowledge support by the Heinrich Heine University 
D\"usseldorf.

\section*{Author Contributions}

C.B. performed the numerical simulations, E.N. developed the analytical model, A.P. and I.K. contributed to 
the interpretation of the results. All authors reviewed the manuscript. 

\section*{Additional Information}
\subsection*{Competing Interests}
The authors declare no competing interests.

\end{document}